# A possible approach to relativistic thermodynamics

Bernhard Rothenstein "Politehnica" University of Timisoara,
Physics Department, Timisoara, Romania
George J. Spix  BSEE Illinois Institute of Technology USA

***Abstract.*** *Considering a system of non-interacting particles characterized by the number N of its constituents and by its Kelvin temperature T, we reduce the transformation of the Kelvin temperature to the transformation mass.*

Many authors present different approaches to relativistic thermodynamics, with different final results, deriving transformation equations for the not very tangible physical quantities: temperature and heat.[1,2,3]. They also mention the present status of the problem. The purpose of our note is to derive transformation equations for temperature and heat, reducing the problem to the transformation of internal energy, a physical quantity proportional to temperature via a relativistic invariant factor.

Consider an ideal monatomic gas consisting of $N$ identical and non-interacting molecules at rest in the K'(X'O'Y') inertial reference frame. The K' reference frame moves with constant velocity $V$ relative to the K(XOY) reference frame, in the positive direction of the common OX(O'X') axes. The axes of the two frames are parallel to each other and at the origin of time in the two frames the origins O and O' are located at the same point in space. In order to characterize the studied system of molecules, observers from K' measure the proper values of the physical quantities introduced, namely its proper temperature $T_0$ and the proper kinetic internal energy $U_0$ given by

$$U_0 = \frac{3k}{2} N T_0 \qquad (1)$$

where *k* represents Boltzmann's constant. By definition $\frac{3k}{2}N$ is a relativistic invariant. The supplementary mass of the gas associated with its internal energy is

$$(\Delta m)_0 = \frac{U_0}{c^2} = \frac{3k}{2c^2} NT_0 .  \qquad (2)$$

The velocity dependence of energy (mass) suggests that detected from K the supplementary mass associated with internal energy is

$$(\Delta m) = \frac{(\Delta m)_0}{\sqrt{1-\frac{V^2}{c^2}}} \qquad (3)$$

where

$$(\Delta m) = \frac{3k}{2c^2} NT \qquad (4)$$

*T* represents the temperature measured in the K reference frame. The result is that temperature should transform as

$$T = \frac{T_0}{\sqrt{1-\frac{V^2}{c^2}}} . \qquad (5)$$

The relativistic invariance of entropy invoked by authors deriving transformation equations for temperature and heat (*Q* in K and $Q_0$ in K') imposes the condition that heat transforms as temperature does i.e.

$$Q = \frac{Q_0}{\sqrt{1-\frac{V^2}{c^2}}} . \qquad (6)$$

Now consider that the ideal gas studied moves with velocity $u_x$ and with velocity $u'_x$ relative to K' related by

$$u_x = \frac{u'_x + V}{1+\frac{Vu'_x}{c^2}} . \qquad (7)$$

Its momentum associated with the motion of the supplementary mass is in K'

$$p'_x = (\Delta m)' u'_x = \frac{U'}{c^2} u'_x . \qquad (8)$$

and

$$p_x = (\Delta m)u_x = \frac{U}{c^2}u_x \qquad (9)$$

in K.

Start with the obvious identities

$$\frac{1}{\sqrt{1-\frac{u_x^2}{c^2}}} = \frac{1+\frac{V}{c^2}u_x'}{\sqrt{1-\frac{u_x'^2}{c^2}}\sqrt{1-\frac{V^2}{c^2}}} \qquad (10)$$

and

$$\frac{u_x}{\sqrt{1-\frac{u_x^2}{c^2}}} = \frac{u_x'+V}{\sqrt{1-\frac{u_x'^2}{c^2}}\sqrt{1-\frac{V^2}{c^2}}}. \qquad (11)$$

Multiplying both sides of (10) by $T_0$ and taking into account the notations introduced above, we obtain

$$T = T'\frac{1+\frac{V}{c^2}u_x'}{\sqrt{1-\frac{V^2}{c^2}}} = \frac{T'+\frac{V}{c^2}u_x'T'}{\sqrt{1-\frac{V^2}{c^2}}} \qquad (12)$$

Multiplying both sides of (11) with $T_0$ and taking again into account the notations used so far we obtain

$$u_x T = \frac{u_x'T'+VT'}{\sqrt{1-\frac{V^2}{c^2}}}. \qquad (13)$$

Equations (12) and (13) tell us that $(T, u_x T)$ and $(T', u_x' T')$ are the components of a "two vector" in one single space dimension approach, the components of which are the temperature and the temperature momentum, $u_x T$ in K and $u_x' T'$ in K'.

The approach presented above could be applied to all scalar physical quantities to which we can associate a proper value say $\Phi_0$ measured in the rest frame of the body, presenting velocity dependence as $\Phi = \frac{\Phi_0}{\sqrt{1-\frac{V^2}{c^2}}}$. $\Phi$ can represent proper time, proper mass, proper (invariant) electric charge or proper electric field intensity, which are the basic physical quantities used in order to build a system of physical units. A procedure similar to that followed above leads to transformation equations for time-space coordinates, mass energy- momentum, electric field intensity-magnetic induction.

Introductory physics textbooks presently in use do not treat the relativistic aspects of thermodynamics. Our approach offers an easy way to relativistic thermodynamics for the introductory physics course.

bernhard_rothenstein@yahoo.com
gjspix@msn.com